\documentclass[conference]{IEEEtran}
\IEEEoverridecommandlockouts
\UseRawInputEncoding
\usepackage{comment}
\usepackage{cite}
\usepackage{amsmath,amssymb,amsfonts}
\usepackage{algorithmic}
\usepackage{graphicx}
\usepackage{textcomp}
\usepackage{balance}
\usepackage{xcolor}
\usepackage{makecell}
\usepackage{booktabs}
\usepackage{multirow}
\usepackage{ragged2e}
\usepackage{subcaption}
\usepackage{makecell}
\usepackage{stfloats}
\usepackage{flushend}
\usepackage{placeins}
\def\BibTeX{{\rm B\kern-.05em{\sc i\kern-.025em b}\kern-.08em
    T\kern-.1667em\lower.7ex\hbox{E}\kern-.125emX}}
\usepackage{tikz}
\usetikzlibrary{positioning, calc}

\begin{document}

\title{EEG-Based Imagined Speech Decoding Using a Hybrid CNN-SNN Architecture
}
\author{\IEEEauthorblockN{
Fatima Shalhoub\IEEEauthorrefmark{1},
Mariam Al Mawla\IEEEauthorrefmark{1}\IEEEauthorrefmark{2},
Kabalan Chaccour\IEEEauthorrefmark{1}\IEEEauthorrefmark{3},~\IEEEmembership{Senior Member,~IEEE}, 
Iv\'an L\'opez-Espejo\IEEEauthorrefmark{4}} Hoda Fares\IEEEauthorrefmark{5}\IEEEauthorrefmark{6},~\IEEEmembership{Member,~IEEE}\\ 
\IEEEauthorblockA{
\IEEEauthorrefmark{1} TICKET Lab., Antonine University, Hadat-Baabda, Lebanon}
\IEEEauthorrefmark{2} Doctoral School of Science, Technology, and Engineering, University of Granada, Granada, Spain\\
\IEEEauthorrefmark{3} University of Technology Belfort-Montb\'eliard, SINERGIES, F-90000 Belfort, France\\
\IEEEauthorrefmark{4} Department of Signal Theory, Telematics and Communications, University of Granada, Granada, Spain\\  
\IEEEauthorrefmark{5} Department of Chemical Engineering, Stanford University, Stanford, USA\\
\IEEEauthorrefmark{6} Department of Electrical and Computer Engineering, Aarhus University, Denmark}
\maketitle

\begin{abstract}

Imagined speech decoding using EEG signals has emerged as a promising frontier in brain-computer interface (BCI) research, particularly to restore communication for individuals with severe speech impairments. However, decoding imagined speech remains a complex task due to the non-stationary, low-amplitude, and highly variable nature of EEG signals. Existing methods often rely on classical machine learning or deep learning models that fail to exploit spike-based temporal dynamics or event-driven firing mechanisms of biological neurons, which are naturally modeled by spiking neural networks (SNNs). In this study, we propose a hybrid decoding pipeline that extracts temporal representations using convolutional neural networks (CNNs) followed by biologically inspired temporal classification via SNNs. To our knowledge, this is the first study to integrate SNNs into EEG-based imagined speech decoding. Experimental results show that the proposed CNN-SNN architecture achieves an accuracy of 80.13\% on the 2020 BCI Competition III benchmark, surpassing existing methods reported in the literature (up to 70.19\%) under comparable evaluation settings. These findings demonstrate the effectiveness of spike-based temporal decoding for imagined speech, highlighting the promise of biologically grounded pipelines for next generation neuromorphic BCI applications.
\end{abstract}


\section{Introduction}
Speech is the natural form of communication between humans. Scientists differentiate two types of speech: 1) \textit{overt speech}, which involves vocal and muscle articulations, and 2) \textit{covert speech}, which refers to inner speech that requires only internal activity of the brain \cite{ref22}.

Overt speech can be adversely affected by a range of neurological disorders that disrupt the neural mechanisms underlying speech production. In particular, cerebrovascular accidents, including cerebellar strokes, are well known to impair motor speech control, resulting in reduced speech clarity and limited verbal communication abilities \cite{ref5}. Nevertheless, cognitive and linguistic functions may remain relatively preserved despite severe motor impairment. Brain-computer interfaces (BCIs) provide a promising assistive technology in decoding covert speech directly from neural activity. The clinical potential of this field has been significantly highlighted by recent advancements in high-bandwidth neural implants which have opened new horizons for speech restoration \cite{chen2024neural}. However, while invasive systems show great promise, there is a parallel need for robust non-invasive solutions. For instance, the authors in \cite{ref2} demonstrated the feasibility of using electroencephalography (EEG) to establish a bidirectional communication channel between the brain and external environments.


The integration of AI into BCI systems has significantly refined the interpretation of complex brain signals, offering new perspectives in healthcare \cite{ref3}. AI-enabled BCIs currently facilitate diagnostic precision and assistive communication for individuals with severe speech impairment  \cite{ref4}. Decoding covert speech is vital for restoring social autonomy in patients suffering from stroke-induced aphasia or other severe neuro-motor impairments \cite{ref35}. However, this process relies on capturing the brain's internal linguistic intent in the absence of overt movement. While imagined speech can be decoded from EEG recordings and converted into text, this task remains a challenging due to the low signal-to-noise ratio and the complex, non-stationary nature of the neural signatures involved.

Existing approaches for EEG-based imagined speech decoding primarily rely on conventional machine learning (ML) and deep neural networks, which operate on continuous-valued activations and do not explicitly model the event-driven dynamics inherent in neural activity. In contrast, spiking neural networks (SNNs) provide a biologically inspired computing paradigm that processes information through discrete spike events over time, thereby introducing an intrinsic temporal dimension that is typically absent in most conventional deep learning (DL) models and more closely mimics the neuronal activity in the brain \cite{khan2025review}. 


This study addresses the problem of classifying imagined speech from EEG data by introducing  a hybrid CNN-SNN architecture.The proposed model marks the first attempt to combine convolutional neural networks (CNN) for effective temporal feature extraction with spiking neural networks (SNNs) for spike-based temporal classification. It is worth noting that SNNs have not yet been explored for EEG-based imagined speech decoding. Moreover, this work demonstrates the potential of spike-driven decision layers to capture the discriminative temporal dynamics of imagined speech signals.


The remainder of this paper is structured as follows: Section \ref{sec:related} reviews work on existing methods for decoding imagined speech using EEG signals. Section \ref{sec:proposed} describes the proposed architecture highlighting the combination of CNN-SNN in imagined speech decoding from EEG signals. Section \ref{sec:results} presents and discusses the obtained results. Finally, Section \ref{sec:conclusion} concludes the paper.

\section{Related Work}
\label{sec:related}

Human language can be represented at multiple linguistic levels, including phonemes, syllables, and words, each associated with subtle and distributed neural activity \cite{ref24}. Imagined speech decoding aims to infer linguistic information directly from brain activity in the absence of overt articulation or acoustic output \cite{ref26}. However, decoding the underlying speech processes remains particularly challenging because speech-related neural patterns are weak, internally generated, and exhibit high inter- and intra-subject variability.

Decoding imagined speech from brain activity typically relies on a multi-stage processing pipeline comprising neural signal acquisition, signal preprocessing, feature extraction, and classification. Neural signals can be acquired using different recording modalities, including electroencephalography (EEG), electrocorticography (ECoG), and magnetoencephalography (MEG), each offering distinct trade-offs between spatial resolution, temporal resolution, and invasiveness \cite{tang2024imagined}. Among these, non-invasive EEG has gained significant attention for imagined speech decoding because of its safety, accessibility, and ease of deployment, despite its limited spatial resolution and susceptibility to noise \cite{ref23}.

EEG-based imagined speech decoding has been explored using a wide range of approaches, spanning classical machine learning methods based on handcrafted features to deep learning models capable of automatically learning spatio-temporal representations from data. Recently, neuromorphic computing paradigms have been proposed as a promising energy-efficient alternative for real-time BCI systems, leveraging event-driven processing and biologically inspired temporal modeling.

\subsection{Classical Machine Learning Methods}
Traditional approaches primarily rely on handcrafted feature extraction followed by conventional classifiers such as support vector machine (SVM), random forest, XGBoost and  k-nearest neighbor (KNN) \cite{ref23}. These methods typically employ time-domain, frequency-domain, time-frequency, or spatial features to capture discriminative characteristics of imagined speech-related neural activity. 

For instance, Reddy and Pachori \cite{Reddy2024MDMD} proposed a multivariate dynamic mode decomposition (MDMD)-based framework to capture coordinated spatiotemporal dynamics across EEG channels, where frequency and power-based features were extracted and classified using a random forest classifier, achieving an accuracy of 73.93\% for short words (``in", ``out", ``up") and 88.9\% for long words (``independent", ``cooperate"). Similarly, Juyal et al. \cite{Juyal2023Connectivity} extracted spatial connectivity features using covariance and phase synchronization index, and applied a parallel ReliefF–neighborhood component analysis (NCA) strategy for EEG channel selection prior to classification with a KNN classifier. Their approach achieved an accuracy of 85.76\% for vowel classification (/a/, /i/, /u/) while reducing the number of EEG channels from 64 to 14–24.

Lopez-Bernal et al. \cite{lopez2023inner} proposed an inner speech EEG decoding framework that extracts time-frequency features using short-time Fourier transform (STFT) and Morlet wavelet transform, along with phase-consistency features based on inter-trial coherence (ITC). These features were classified using SVM and KNN classifiers, achieving accuracies of 58.75\% and 67.5\%, respectively, for identifying four directional Spanish words (``up", ``down", ``right",  ``left").

\subsection{Deep Learning Methods}
Despite these encouraging results, the reliance of classical ML methods on handcrafted features limits their scalability and robustness. Consequently, deep learning methods have been widely adopted for imagined speech decoding from EEG signals due to their ability to automatically learn both temporal and spatial feature representations from neural signals. 

Leerskov et al. \cite{ref9} demonstrated the effectiveness of CNNs, reporting a 13.2\% improvement in accuracy compared to KNN when classifying the words `pet' and `naw' from Kent's phoneme list. They applied fast Fourier transform (FFT) for feature extraction and cross covariation between frequency channels before CNN classification. More recent approaches include the Diff-E proposed by Kim et al. \cite{kim23g_interspeech}, which integrates denoising diffusion probabilistic models (DDPMs) with conditional autoencoders for EEG decoding. This approach achieved an accuracy of 60.63\% $\pm$ 8.03\% and an AUC of 90.39\% $\pm$ 3.48\%, significantly outperforming DeepConvNet (21.00\% $\pm$ 4.27\% accuracy, 69.55\% $\pm$ 3.51\% AUC).

Moreover, Alharbi and Alotaibi \cite{Alharbi2024Hybrid} proposed a hybrid DL framework that transforms EEG signals into sequential topographic brain maps and applies a combination of 3D CNNs and recurrent neural networks (RNNs) to capture spatio-temporal features, achieving an average classification accuracy of 77.8\% across multiple imagined words. Agarwal and Kumar demonstrated in \cite{Agarwal2022LSTM} an imagined speech recognition system based on long-term memory (LSTM) using 32-channel EEG recordings of four words/phrases (``sos", ``stop", ``medicine", ``washroom", ``come-here"), reporting an accuracy of 73.6\%.

To address the challenge of low spatial EEG resolution,  Li et al. \cite{ref15} introduced the Large Kernel-ConvMixer architecture which combines extended receptive field kernels with residual networks to improve spatial and temporal feature extraction. This approach achieved an accuracy of 75.73\%, outperforming the previous best-performing EEGNet transformer combination while also reducing training time and parameter count.

\subsection{Neuromorphic Methods} 
Although DL architectures dominate current imagined EEG-based speech decoding pipelines, they remain computationally demanding, posing challenges for low-latency, wearable, and energy-efficient BCI deployment. Neuromorphic models, represented by spiking neural networks (SNNs), offer a brain-inspired  alternative by representing the information as sparse spike events and explicitly modeling neural temporal dynamics. Such properties have motivated their adoption to BCI systems, particularly applications requiring real-time processing and on-device inference. This has been demonstrated in recent studies, for instance,  Corradi et al. \cite{ref34} explored the feasibility of neuromorphic architectures for decoding complex cognitive states with competitive accuracy and low computational cost.


Although classical ML and DL methods have shown promising performance in decoding imagined speech, they are often computationally demanding and inefficient for real-time applications. In contrast, neuromorphic computing, while not yet explored for imagined speech decoding, offers inherent advantages through low-power and event-driven processing and temporal modeling. Table \ref{tab:decoding_methods} summarizes common advantages and limitations of existing imagined speech decoding methods compared to neuromorphic computing-based ones.

\begin{table}[h]
\caption{Advantages and limitations of imagined speech decoding methods and neuromorphic approaches}
\label{tab:decoding_methods}
\RaggedRight
\centering
\begin{tabular}
{p{2cm} p{2.8cm} p{2.8cm}}
\toprule
\textbf{Method} & \textbf{Advantages} & \textbf{Limitations} \\
\toprule
\textbf{Classical ML} & Simple and interpretable; fast training time & Requires manual feature engineering; limited scalability\\
\midrule
\textbf{Deep Learning} & Learns features automatically; high accuracy & Needs large datasets; computationally demanding \\
\midrule
\textbf{Neuromorphic} & Biologically inspired processing; low power; real-time processing capability & Limited research; hardware not widely available \\
\bottomrule
\end{tabular}
\end{table}

\section{Proposed Architecture}
\label{sec:proposed}
\begin{figure*}[!t]
\centerline{\includegraphics[scale=0.43]{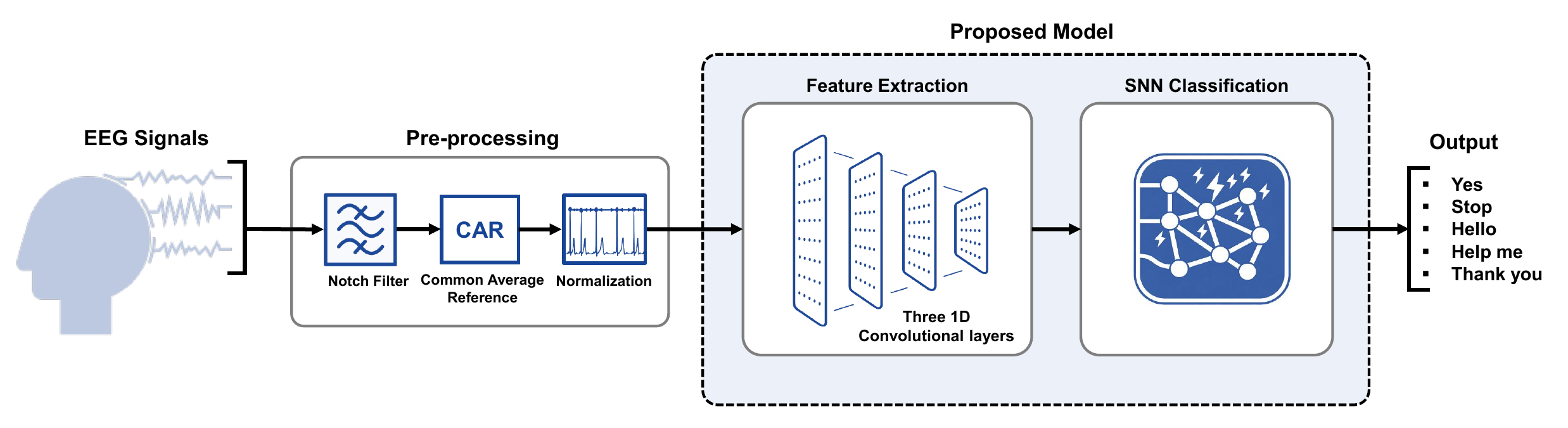}}
\caption{Proposed pipeline for EEG-based imagined speech decoding.}
\label{EEG_Processing}
\end{figure*}

Motivated by the limitations of classical and deep learning approaches, this paper proposes a neuromorphic inspired architecture for EEG-Based imagined speech. It adopts a hybrid CNN-SNN model for EEG-based imagined speech classification, as illustrated in Fig. \ref{EEG_Processing}. This combination allows us to effectively extract temporal EEG features while leveraging biologically inspired temporal processing. It consists of a convolutional feature extraction module followed by a temporal SNN classifier based on leaky integrate-and-fire (LIF) neurons.
\subsection{EEG Dataset}
\label{sec:eeg_dataset}
The 2020 BCI Competition III dataset \cite{bci_competition_committee_2025} is a publicly available multi-class non-invasive EEG database captured during imagined speech tasks. The dataset includes EEG data from 15 healthy participants who were instructed to silently imagine speaking five distinct words or phrases (``Hello'', ``Help me'', ``Stop'', ``Thank you'', and ``Yes'') across multiple trials. The brain signals were recorded using a 64-channel EEG system aligned to the international 10-20 electrode placement standard \cite{acharya_hani_cheek_thirumala_tsuchida_2016}, ensuring full-scalp spatial coverage. Recordings were sampled at 256 Hz. Each participant completed 80 trials per class, resulting in multi-channel time-series EEG segments. 
The actual dataset provides predefined training, validation, and test subsets comprising 60, 10, and 10 trials per class, respectively.  As the labels for the test set are not available, only the provided training and validation subsets are used in this study as follows:

\begin{itemize}
\item \textbf{Training set (85\%):} 4500 labeled trials from the training subset.
\item \textbf{Test set (15\%):} 750 labeled trials from the validation subset for prediction and model evaluation.
\end{itemize}

\subsection{EEG Pre-processing}
Prior to feature extraction and classification, the raw EEG signals are processed using a standardized pre-processing pipeline designed to reduce noise and inter-channel variability. First, a notch filter centered at 50 Hz is applied to all EEG trials to suppress power-line interference. The filter is implemented as a second-order infinite impulse response (IIR) with a quality factor $Q=35$ and a sampling frequency of 256 Hz. Zero-phase forward-backward filtering is employed to prevent phase distortion and filtering is performed independently along the temporal dimension of each EEG channel. Following notch filtering, a common average reference (CAR) spatial filter is applied on a per-trial basis. At each time point, the mean signal across all channels is subtracted from each channel, thereby reducing common noise components. Subsequently, the EEG signals are normalized using $z$-score standardization to reduce amplitude variability across trials and time within each EEG channel. For multi-channel time-series data, normalization is performed independently for each channel by computing the mean and standard deviation over all time samples in the training set, which are then applied to the test data to ensure consistent scaling and prevent data leakage.

\subsection{Model Building}
This section details the proposed classification pipeline, which consists of two processing sequential stages. First, a convolutional feature extraction module that learns compact temporal representations from the preprocessed EEG signals. These representations are then processed by a time-stepped SNN classifier that performs spike-based computation and temporal integration across discrete time steps. The resulting spiking activity is decoded using a spike rate–based scheme for class prediction. The implementation details of each module are described in the following subsections.

\subsubsection{CNN Feature Extractor}
The CNN module is designed to extract discriminative temporal features from multi-channel EEG signals using one-dimensional temporal convolutions, as illustrated in Fig.~\ref{fig:model}(a). This CNN receives EEG segments represented as a multi-channel time-series tensor [\texttt{batch} $\times$ \texttt{channels} $\times$ \texttt{time}], where 1D convolutions are applied along the temporal dimension to capture complex temporal dependencies within each EEG channel independently while preserving the spatial structure across channels.
The first layer applies 256 temporal filters with a kernel size of 11, enabling the capture of short- and mid-range temporal patterns in the signals while also preserving the temporal resolution through symmetric padding. This is followed by batch normalization to stabilize feature distributions, ReLU activation to introduce nonlinearity, dropout to mitigate overfitting, and average pooling with a factor of 2 to reduce dimensionality and noise sensitivity. The second 1D convolutional layer applies a bottleneck expansion (kernel size of 1) with 320 channels to enhance representational capacity with limited computational overhead. The resulting features are subsequently processed by a third convolutional layer with 64 output channels, kernel size of 9, and dilation factor of 2 to expand the temporal receptive field without increasing parameter count. The final convolution is followed by batch normalization, ReLU activation, and dropout, along with average pooling using a pooling factor of 4. The output of these CNN layers is temporal feature representations. They are then integrated with SNN layers for deeper temporal processing and final decision making.

\subsubsection{SNN Classifier}
SNNs are considered the third generation of neural network models and are inspired by the event-driven communication and temporal processing mechanisms of biological neural systems \cite{maass1997networks}. Unlike conventional artificial neural networks (ANNs), which propagate continuous activation values, SNNs transmit information via discrete spike events over time, enabling efficient modeling of temporal dependencies.

The sequential feature embeddings produced by the convolutional module are first converted into spike trains using a temporal encoding strategy, where temporal variations in the CNN feature representations are captured through spike timing. These encoded spike trains are provided to the SNN across discrete time steps.

All spiking neurons in the network are modeled using LIF dynamics. Each LIF neuron integrates the incoming weighted spike currents into a membrane potential $v(t)$, which evolves according to leaky dynamics governed by a decay factor $\beta\in(0,1)$. The decay factor $\beta$ regulates the neurons' temporal memory and sensitivity to past inputs. When the membrane potential exceeds a predefined firing threshold $\theta$, the neuron emits a spike and resets its membrane potential, mimicking biological neuronal firing \cite{ref28, ref27}.

\begin{figure*}[!t]
    \centering
    \begin{subfigure}{0.99\linewidth}
        \centering
        \includegraphics[scale=0.78]{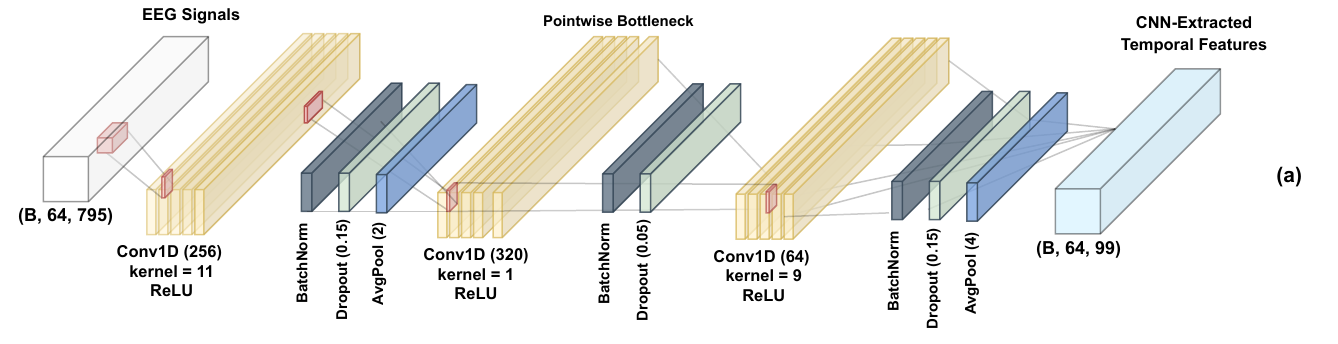}
        \label{fig:cnn}
    \end{subfigure}

    \vspace{-0.5em} 

    \begin{subfigure}{0.99\linewidth}
        \centering
        \includegraphics[scale=0.78]{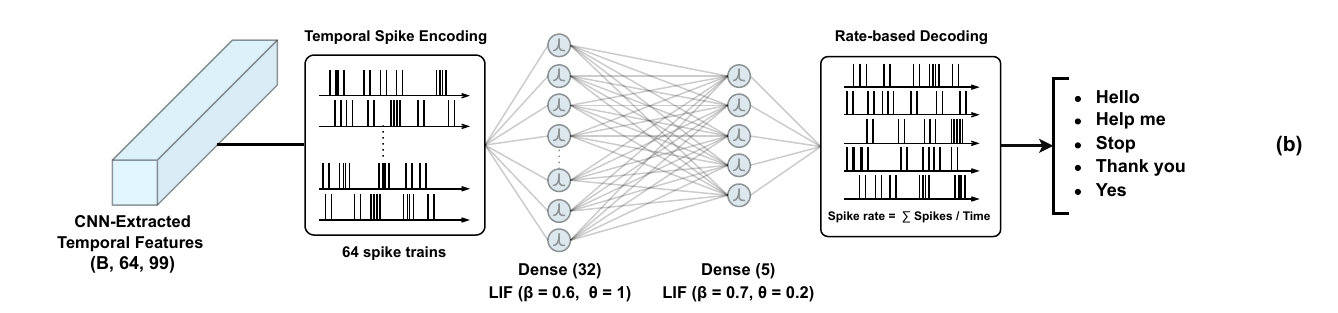}
        \label{fig:snn}
    \end{subfigure}

    \caption {Architecture of the proposed CNN-SNN model illustrating (a) the CNN-based feature extraction module and (b) the SNN-based imagined speech classification module.}
    \label{fig:model}
\end{figure*}

The SNN module consists of two fully-connected spiking layers, as illustrated in Fig.~\ref{fig:model}(b). The first layer maps the 64 CNN-derived spike trains to 32 hidden spiking neurons using LIF dynamics with a decay factor of $\beta=0.6$ while keeping the default firing threshold $\theta=1.0$. This configuration allows neurons to selectively respond rapidly to highly-synchronized temporal variations within the CNN features while filtering out stochastic noise. The second spiking layer projects the hidden spike activity to 5 output neurons, each corresponding to an imagined speech class. This layer uses a slower decay ($\beta=0.7$) and a reduced threshold ($\theta=0.2$) to facilitate more stable temporal accumulation and controlled spike generation at the decision level.

The network operates in a time-stepped manner. At each discrete time step, membrane potentials evolve dynamically based on incoming spikes and neuron dynamics \cite{kugele2020efficient}. Through stacked spiking layers, the network learns temporal patterns by accumulating and propagating spike activity over time, allowing it to model temporal dependencies in the CNN feature sequences. 

The output spikes of the final spiking layer are then accumulated and averaged over time to obtain the mean firing rate $r_k$ for each output neuron $k$:
\begin{equation}
    r_k = \frac{1}{T} \sum_{t=1}^{T} s_k(t),
\end{equation}
where $s_k(t)\in\{0,1\}$ indicates whether the class-$k$ output neuron fired at time $t$. To obtain the final class prediction, the predicted class $\hat{k}$ is determined using a rate-based decoding scheme by selecting the output neuron with the highest mean firing rate:
\begin{equation}
    \hat{k} = \arg\max_k r_k,
\end{equation}
where $k \in\{1,2,3,4,5\}$ denotes the possible imagined speech classes.

To address the non-differentiability of spike generation, the SNN is trained using surrogate-gradient backpropagation based on an arc-tangent approximation, enabling stable gradient flow through the spiking layers. This spike-based temporal integration allows the SNN to capture dynamic patterns in the learned feature sequences while maintaining sparse, event-driven computation, providing an efficient and biologically inspired alternative to conventional ANN-based classifiers.

\section{Results and Discussion}
\label{sec:results}
All experiments were conducted on the 2020 BCI Competition III dataset introduced in Subsection~\ref{sec:eeg_dataset} using a subject-dependent supervised classification setup. The provided dataset split was used, with 85\% of the trials for training and 15\% for testing. Performance was evaluated using classification accuracy and F1-score.

The proposed CNN-SNN model was trained for 350 epochs using the AdamW optimizer with decoupled weight decay. CNN layers were optimized with a learning rate of $3 \times 10^{-4}$, while the SNN classifier was trained using a smaller learning rate of $5 \times 10^{-5}$ to ensure stable membrane potential evolution and spike generation. A weight decay of $2 \times 10^{-4}$ was applied only to high-dimensional convolutional and fully-connected weights to mitigate overfitting while avoiding disruption of the spiking neuron dynamics. The model was optimized using the categorical cross-entropy loss, computed between the predicted class scores (highest mean firing rates) and the ground-truth labels. The proposed CNN-SNN model achieved an average classification accuracy of 80.13\% and F1-score of 80.14\% on the 2020 BCI Competition III dataset.

\begin{table}[t!]
\caption{Comparison of classification accuracies (in \%) across selected EEG-based imagined speech decoding systems on the 2020 BCI Competition III dataset}
\label{tab:comparison_BCI}
\RaggedRight
\centering
\def\arraystretch{1.2}
\begin{tabular}
{p{2.2cm} p{3.6cm} >{\centering\arraybackslash}p{1.75cm}}
\toprule
\textbf{Study} & \textbf{Architecture} & \textbf{Accuracy (\%)} \\
\toprule
Lee et al.~\cite{9481777} & Spectral Entropy + Siamese Neural Network + KNN & 48.10 \\ 
Ko et al.~\cite{10.1007/978-3-031-02444-3_25} & Spectro-Spatio-Temporal CNN & 70.19 \\ 
Zheng et al.~\cite{ZHENG2023105324} & Statistical Features + Random Forest & 58.51 \\ 
Rousis et al.~\cite{10715364} & EEGNet-SPDNet & 66.93 \\ 
Li et al.~\cite{10782407} & Self-attention Module + Transfer Learning & 69.00 \\ 
\midrule
\textbf{Proposed Model} & \textbf{CNN-SNN} & \textbf{80.13} \\
\bottomrule
\end{tabular}
\end{table}

To evaluate the effectiveness of the proposed model, we compare its performance with previously reported state-of-the-art methods on the 2020 BCI Competition III dataset. The selected studies include both conventional ML and DL models that report classification accuracy under comparable subject-dependent evaluation settings. Table ~\ref{tab:comparison_BCI} shows that the proposed CNN-SNN model outperforms existing approaches, achieving the highest classification accuracy (80.13\%). Unlike prior approaches that rely exclusively on conventional CNNs ~\cite{10.1007/978-3-031-02444-3_25} or attention mechanisms ~\cite{10782407}, our proposed model leverages temporal spike-based dynamics for classification. This result highlights the importance of integrating biologically inspired spiking neurons for imagined speech decoding.

In addition to the classification accuracy, computational efficiency was evaluated. Inference time was measured in evaluation mode on a Google Colab NVIDIA Tesla T4 GPU, achieving an average prediction time of approximately 98 $\pm$ 3 ms per trial. Given that each EEG trial corresponds to a 2-second segment, the latency introduced by the model is small relative to the signal duration, suggesting potential for real-time BCI applications ($\sim$0.05 real-time factor). Although the model exhibits low inference latency, this study was conducted on pre-recorded EEG data in an offline setting, and full validation under real-time conditions remains necessary.

\section{Conclusion}
\label{sec:conclusion}
This work introduces a hybrid CNN-SNN architecture for EEG-based imagined speech decoding, marking the first use of SNNs in this task, which integrates deep temporal feature extraction with biologically inspired spiking dynamics. The convolutional block captured temporal patterns from the signals, while the SNN classifier performed temporal and sparse event-based processing for efficient classification. The proposed model achieved an average classification accuracy above 80\% on the 2020 BCI Competition III dataset, demonstrating the efficacy of controlled spiking dynamics combined with CNN-extracted features. The discussion showed that the proposed CNN-SNN model achieved the highest classification accuracy compared to recent CNN-based approaches applied on the same dataset. These results highlight the potential of SNNs for neural decoding, especially in real-time and power-constrained applications. Future work will investigate the model’s performance on larger, multi-subject datasets to address inter-subject variability in non-invasive EEG. 


\section*{Acknowledgment}
This work was supported by the Spanish Ministry of Science and Innovation under the ``Ram\'on y Cajal'' programme (RYC2022-036755-I), and by the Lundbeck Foundation in Copenhagen, Denmark (R402-2022-1413).

\balance
\bibliographystyle{IEEEtran}
\bibliography{bibtex}
\clearpage
\end{document}